\documentclass[useAMS,usenatbib]{mn2e}

\usepackage{graphicx}
\usepackage{natbib}

\title[PL$_K$ relation for Galactic Globular Clusters]
{The RR Lyrae Period - K Luminosity relation for Globular Clusters: an 
observational approach\thanks{Based on 
observations collected at the European Southern Observatory
within the observing programs 49.5-0021, 51.5-0024, 59.E-0340, 64.N-0038, 
68.D-0287 and at the Telescopio Nazionale Galileo.}}
\author[A. Sollima, C. Cacciari and E. Valenti]{A. Sollima$^{1}$\thanks{E-mail:
antonio.sollima@bo.astro.it (AS)}, C. Cacciari$^{2}$ and E. Valenti$^{2}$\\
$^{1}$Dipartimento di Astronomia, Universit\`a di Bologna, via Ranzani 1,
Bologna, 40127-I, Italy\\
$^{2}$INAF Osservatorio Astronomico di Bologna, via Ranzani 1,
Bologna, 40127-I, Italy}
\begin{document}

\date{Accepted 2006, ???; Received 2006, ???; in original form
2006, ???}

\pagerange{\pageref{firstpage}--\pageref{lastpage}} \pubyear{2006}

\maketitle

\label{firstpage}

\begin{abstract}
The Period - metallicity - K band luminosity (PL$_K$) 
relation for RR Lyrae stars in 15 Galactic globular clusters and in the LMC
globular cluster Reticulum has been derived. It is based on accurate near 
infrared (K) photometry combined with 2MASS and other literature data.
The PL$_K$ relation has been calibrated and compared with the previous empirical and 
theoretical determinations in literature. The zero point of the absolute 
calibration has been obtained from the K magnitude of RR Lyr whose distance 
modulus has been measured via trigonometric parallax with HST. 
Using this relation we obtain a distance modulus to the LMC of 
$(m-M)_0 = 18.54 \pm 0.15$ mag, in good agreement with recent 
determinations based on the analysis of Cepheid variable stars. 
\end{abstract}


\section{Introduction}

RR Lyrae, as well as classical Cepheids, are considered standard candles for 
estimating stellar distances in the Milky Way and to Local Group galaxies.
They are produced in old stellar 
populations, and hence they provide important information for an understanding 
of the age, structure and formation of their parent stellar systems. 
As typical Population II stars, RR Lyrae are abundant in globular clusters 
and have been the subject of a huge number of studies for more than a century.

Although the pulsation theory explains quite well the connection 
between most of the involved physical quantities (van Albada \& Baker 1971; 
Caputo, Marconi \& Santolamazza 1998), the dependence of intrinsic luminosity on
metal content has been the subject of debate for nearly three 
decades (see Smith 1995 for a review). 
Only recently the slope of the luminosity-metallicity relation 
M$_V$-[Fe/H] seems to be converging towards a value $\sim$0.20-0.23 that 
appears to be ``universal'' as supported by the most accurate studies of 
field RR Lyrae stars in the Milky Way (Fernley et al. 1998; Chaboyer 1999) 
and in the Large Magellanic Cloud (LMC, Gratton et al. 2004), and globular 
clusters in M31 (Rich et al. 2005). 

Near infrared observations of variable stars present several advantages over optical
investigations: a smaller dependence on interstellar extinction and metallicity, a smaller
pulsational amplitude and more symmetrical light curves, and hence good mean magnitudes.   
In the last 20 years the calibration of the RR Lyrae period-IR luminosity
relation (PL$_K$) has been the subject of several empirical and theoretical
investigations (see Sect. \ref{method}), 
but there still remains some degree of uncertainty on the 
dependence of $M_{K}$ on period and metallicity.
To solve the problem of the dependence of $M_{K}$(RR) on these physical quantities, 
a large sample of RR Lyrae stars is needed, spanning a wide range of [Fe/H], for which 
accurate K and [Fe/H] measurements are available.

In this paper we present an accurate analysis of the infrared photometric
properties of 538 RR Lyrae variables (376 RRab and 162 RRc) in 16
globular clusters (GC) with $-2.15<[Fe/H]<-0.9$. This more than doubles
the number of stars used in the previous largest study of this type (Nemec, 
Linnell Nemec \& Lutz 1994). By means
of such a large database we calibrated the PL$_K$ relation constraining its
dependence on period and metallicity on a strictly observational basis.

In \S 2 we describe the sample of RR Lyrae stars and the used photometric IR datasets.
\S 3 is devoted to the description of the adopted method to calibrate the PL$_K$ relation. 
In \S 4 we use the derived PL$_K$ relation to estimate the distance to the 
calibrator GCs and to a sample of field RR Lyrae stars in the LMC, and compare the results
with the previous determinations in literature. 
Finally, we summarize our results in \S 5. 

\section{Observations}

The data used in the present analysis consist of K photometry for the central 
regions of 9 GCs
(M4, M5, M15, M55, M68, M92, M107 and $\omega$ Cen) derived from a set of images secured 
at the Telescopio Nazionale Galileo (TNG, Canary Islands), using the near-IR 
cameras ARNICA, and at the European Southern Observatory (ESO, La Silla), using 
the near-IR camera IRAC-2 and SOFI.
A detailed description of the data reduction and calibration procedure
can be found in Ferraro et al. (2000), Valenti et al. (2004), Valenti, Ferraro \& Origlia
(2004) and Sollima
et al. (2004).
All measured instrumental magnitudes were transformed into 
the Two-Micron All-Sky Survey (2MASS)\footnote{See 
Skrutskie et al. (2006), Explanatory Supplement to the 2MASS All Sky Data Relase,
http://www.irsa.ipac.caltech.edu/cgi-bin/Gator/} photometric system.

To extend our analysis to the outer regions of these clusters, we
correlated our catalogs with the database obtained by 2MASS that extends to
a wide area up to 15' from the cluster centers. 
The 2MASS K photometry for 5 additional GCs (M22, NGC 3201, NGC 5897, NGC 6362 and NGC 6584) 
was also considered.

For each cluster we identified a large number of variable stars by 
cross-correlating the IR catalog with the most comprehensive catalog 
of GC variable stars available in literature (Clement et al. 2001). 
Since the adopted K magnitudes are the average of repeated exposures and the K light 
curves of RR Lyrae variables have a fairly sinusoidal low amplitude shape, we
assumed our K magnitudes as the mean magnitudes of the identified variables. 
Note that the individual K values from different datasets generally agree within 
less than 0.1 mag. So, we assumed $\pm$0.1 mag as a plausible error to attach 
to the mean K magnitudes used in the following analysis.

In addition, we considered the K photometry of RR Lyrae stars taken by 
Longmore et al. (1990, for variables in the clusters M3, M4, M5, M15, M107 and 
NGC 3201), Storm et al. (2004, IC4499), Butler (2003, M3), Dall'Ora et al.
(2004, Reticulum) and 
Del Principe et al. (2005, M92). All magnitudes were reported to the homogeneous
photometric system of 2MASS using the transformation equations provided by
Carpenter (2001).   

Table 1 lists for each calibrator GC the metallicity 
(in the Carretta \& Gratton 1997 scale) and the reddening coefficient E(B-V)
from Ferraro et al. (1999) together with the number of RR Lyrae considered in the 
present analysis.

\begin{table}
\label{summ}
 \centering
  \caption{The sample of calibrator GCs. Metallicities are in the CG scale. }
  \begin{tabular}{@{}lcccr@{}}
  \hline
   Name     & [Fe/H] & E(B-V) & \# RR Lyrae  & Ref. \\
 \hline
 M3           & -1.34 & 0.01  & 28 & L90, B03\\
 M4           & -1.19 & 0.36  & 31 & L90, F00, 2M\\
 M5           & -1.10 & 0.035 & 86 & L90, V04, 2M\\
 M15          & -2.15 & 0.09  & 52 & L90, VFO04, 2M\\
 M22          & -1.53 & 0.38  & 17 & 2M\\
 M55          & -1.61 & 0.07  &  8 & F00, 2M\\
 M68          & -1.95 & 0.04  & 23 & F00, 2M\\
 M92          & -2.15 & 0.025 & 11 & D05, 2M\\
 M107         & -0.87 & 0.33  & 22 & L90, F00, 2M\\
 IC4499       & -1.26 & 0.24  & 24 & St04\\
 NGC 3201     & -1.23 & 0.21  & 60 & L90, 2M\\
 NGC 5897     & -1.59 & 0.08  &  5 & 2M\\
 NGC 6362     & -0.92 & 0.08  & 31 & 2M\\
 NGC 6584     & -1.30 & 0.11  & 32 & 2M\\
 $\omega$ Cen & -1.60$^2$ & 0.11 & 82 & L90, So04, 2M\\
 Reticulum   & -1.48 & 0.03   & 32 & D04\\
 \hline
\end{tabular}
$^2$ Because of the well known metallicity spread
 among RR Lyrae stars in this cluster (Sollima et al. 2006 and reference
therein), we took into account only the metal-poor ($[Fe/H]<-1.4$ )
$$ $$
References: L90: Longmore et al. (1990); 
2M: 2MASS Skrutskie et al. (2006); F00: Ferraro et al. (2000); B03: Butler (2003); 
V04: Valenti et al. (2004); VFO04: Valenti, Ferraro \& Origlia (2004); St04:
Storm (2004); So04: Sollima et al. (2004); D04: Dall'Ora et al. (2004); D05: Del Principe
et al. (2005).  
\end{table}

\section{Method}
\label{method}

Theoretical studies on RR Lyrae stars performed in the past indicated a
relationship between (infrared) luminosity, metallicity and period of the form
\begin{equation}
\label{eq1}
M_{K}=\alpha~log P_{F}+\beta~[Fe/H]+\gamma
\end{equation}
where $P_{F}$ indicates the period of the variables pulsating in the fundamental mode 
(RRab-type). 
The apparent K magnitude can be obtained by adding to both sides of eq (\ref{eq1}) 
the K distance modulus
\begin{equation}
\label{eq2}
K=\alpha~log P_{F}+\beta~[Fe/H]+\gamma+(m-M)_{K}
\end{equation}

Several authors have calibrated PL$_K$ relations, in the form
of eq. (\ref{eq1}), on the basis
of observations in GCs and field RR Lyrae stars. In Table 2 the
obtained empirical and theoretical PL$_K$ relations are summarized. All the empirical 
relations use metallicities in the Zinn \& West (1984) metallicity scale. 
Note that while the dependence on the 
period is quite consistent between the empirical and the theoretical values, 
the dependence on metallicity is about 3 times larger in theoretical estimates. 

In the following sections we describe the adopted method to estimate the 
coefficients $\alpha, \beta$ and $\gamma$ of eq. (\ref{eq1}) using only
observational constraints.

\begin{table}
\label{past}
 \centering
  \caption{Determinations of PL$_K$ relations from previous studies.}
  \begin{tabular}{@{}lccr@{}}
  \hline
   Ref.     & $\alpha$ & $\beta$  & $\gamma$ \\
 \hline
Empirical                         &        &      \\
\hline
 J88                               & -1.72 &      & -0.74\\
 L90  (based on Jones et al. 1988) & -1.72 & 0.04 & -0.65\\
 L90  (based on Liu \& Janes 1990) & -2.26 & 0.08 & -0.76\\
 LJ90                              & -2.72 &      & -0.99\\
 J92                               & -2.03 & 0.06 & -0.72\\
 J92                               & -2.33 &      & -0.88\\
 C92                               & -2.33 &      & -0.88\\
 S93                               & -2.95 &      & -1.07\\
 N94                               & -2.40 & 0.06 & -1.27\\
 N94                               & -2.40 &      & -0.95\\
 FS98                              & -2.34 &      & -0.88\\
\hline
Theoretical                       &        &      \\
\hline
B01                               & -2.071 & 0.167 & -0.77\\
B03                               & -2.101 & 0.231 & -0.77\\
C04                               & -2.353 & 0.175 & -0.89\\ 
 \hline
\end{tabular}

References: J88: Jones, Carney \& Latham (1988); L90: Longmore et al. (1990); LJ90: Liu \& Janes (1990); 
J92: Jones et al. (1992); C92: Carney, Storm \& Jones (1992); S93: Skillen et al. (1993); 
N94: Nemec et al. (1994); FS98: Frolov \& Samus (1998); 
B01: Bono et al. (2001); B03: Bono et al. (2003); C04: Catelan, Pritzl \& Smith (2004). 
\end{table}

\subsection{Dependence on Period ($\alpha$)}
\label{period}

\begin{figure*}
 \includegraphics[width=12.cm]{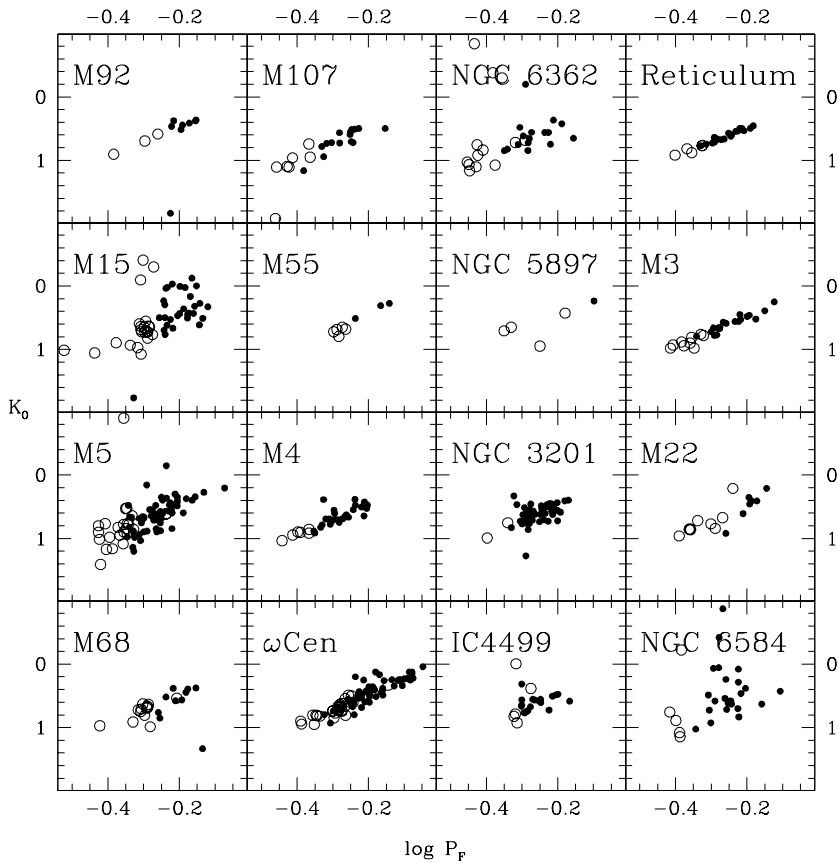}
\caption{PL$_K$ relation for the RR Lyrae of our calibrator GCs. 
Filled circles are the RRab variables, open circles are the RRc variables whose periods have been 
fundamentalized. K magnitude were scaled to the same reference distance and
metallicity (see Sect. \ref{period}).}
\label{all}
\end{figure*}

\begin{figure}
 \includegraphics[width=8.7cm]{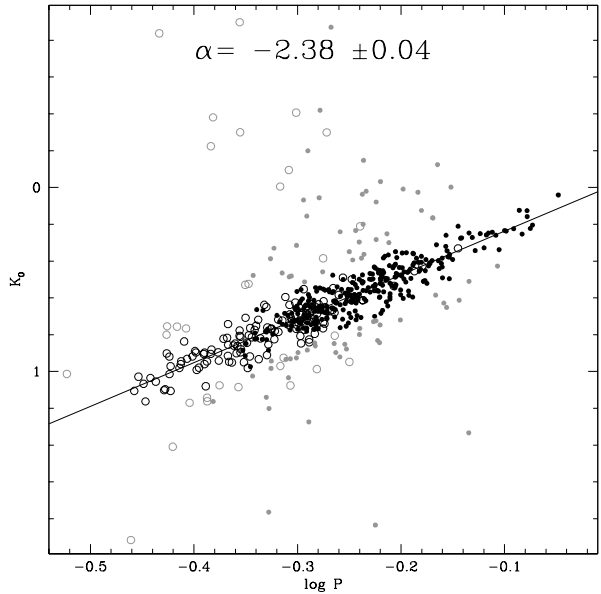}
\caption{PL$_K$ relation for the 538 RR Lyrae of our sample. The solid line 
represent the resulting fit. Filled circles are the 
RRab variables, open circles are the RRc variables whose periods have been 
fundamentalized. K magnitude were scaled to the same reference distance and
metallicity (see Sect. \ref{period}). Grey symbols mark the points rejected to
the fit.}
\label{alpha}
\end{figure}

The advantage of using GCs in constraining the coefficients $\alpha, \beta$ and
$\gamma$ lies in the fact that all the stars in a given cluster are 
at the same distance, and can be considered to share the same metal content and be 
subject to the same extinction effect.  
Therefore, within a given globular cluster eq. (\ref{eq2}) can be written in the form
\begin{equation}
\label{eq3}
K-\alpha~log P_{F}=C$$
\end{equation}
where C contains the information on reddening, metallicity and distance for that
cluster.
As a first step we estimate the constant C for each calibrator GC by
assuming a first guess value $\alpha=-2.4$ (Nemec et al. 1994) 
and averaging the left-hand side of eq. (\ref{eq3}) for all the cluster RR Lyrae stars. 
First overtone RR Lyrae (RRc) are included in the analysis after correcting 
their periods by adding a constant term ($\Delta log~P=0.127$).  
Outliers are identified as those stars whose C values differ from the median 
more than 3 times the semi-interquartile range\footnote{The interquartile range is defined as the distance between the $25^{th}$ and the $75^{th}$ percentile of the parent cumulative distribution of a given set of data.} and rejected. 
Then, we scale the apparent K magnitudes for the corresponding value of C for each
sample of GC RR Lyrae, merge all the samples together and best-fit eq. (\ref{eq3}) 
by ordinary least squares thus estimating an updated value of $\alpha$. An
iterative procedure is carried out until a stable value of $\alpha$ is
obtained. In Fig. \ref{all} the distribution of the RR Lyrae scaled K magnitudes
($K_{0}=K-C$) as a function of periods is shown for each of the 16 calibrator GCs. 
The procedure described above converges in few iterations to a global value of
$\alpha=-2.38\pm0.04$ (see Fig. \ref{alpha}) and is insensitive to the first guess of $\alpha$.  
The value of $\alpha$ estimated above is in good agreement with both empirical
and theoretical estimates obtained in previous studies (see Table 2).

\begin{figure}
 \includegraphics[width=8.7cm]{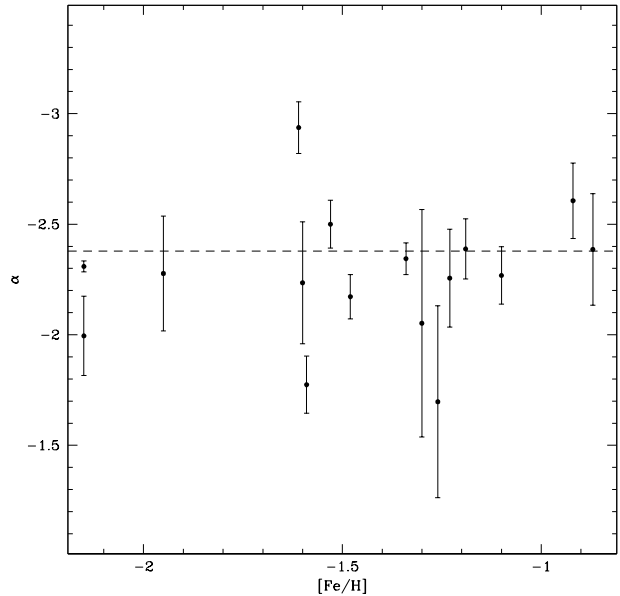}
\caption{The parameter $\alpha=\delta K/\delta logP$ for RR Lyrae stars 
in various globular clusters, as a function of the cluster metallicity. The
dashed line indicate the general value of $\alpha=-2.38$ estimated for the entire
sample.}
\label{pkmet}
\end{figure}

In Fig \ref{pkmet} we plot the individual slopes $\alpha_{i}$ calculated 
separately for each calibrator GC as a function of the cluster metallicity. Note that, 
whereas an average value around $\alpha=-2.35$ is generally a good approximation for most 
clusters, $\alpha$ can significantly vary along the entire metallicity
range. A Spearman rank correlation test gives a 78\% probability that the two
variables ($\alpha$ and $[Fe/H]$) are correlated between them.
This effect is due to the complex interplay of dependences of the
mass-luminosity, luminosity-temperature and K bolometric correction-temperature
relations on metallicity that drives metal-rich variables to steeper PL$_K$
relations.   
So, from this point of view the PL$_K$-[Fe/H] relation is a bit less 
{\em universal} than we might wish.

\subsection{Dependence on Metallicity ($\beta$)}
\label{met}

\begin{figure}
 \includegraphics[width=8.7cm]{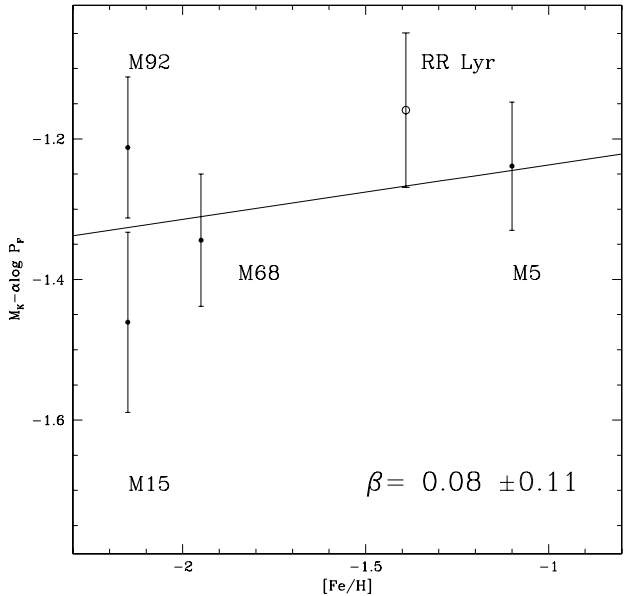}
\caption{Bestfit of eq. (\ref{eq1}) for the four calibrator GCs with known 
distances from Carretta et al. (2000). The open dot indicate the location in the
diagram of the variable RR Lyr.}
\label{beta}
\end{figure}

Once the coefficient $\alpha$ is derived, two other parameters still remain to be
derived in eq. (\ref{eq2}) namely the slope $\beta=\delta K/\delta [Fe/H]$ and the
zero point $\gamma$. Since both metallicity and distance
vary from cluster to cluster, there is an evident degeneracy between the
coefficient $\beta$ and the additive term $\gamma+(m-M)_{K}$. No constraints can
be assessed for $\beta$ without an "a priori" knowledge on the distance and
reddening of a subsample of our calibrator GCs.
In order to disentangle this degeneracy in a homogeneous and completely observational way, we 
consider the distance determinations by Carretta et al. (2000) 
based on the application of the MS fitting technique to a sample of 12 GCs 
using Hipparcos trigonometric parallaxes of 
local subdwarfs. Table 3 lists the adopted
distances and reddening coefficients for the 4 GCs (M5, M15, M68 and M92) 
in common between our sample and that of Carretta et al. (2000). 
Absolute distance moduli were converted to
K band ones adopting the reddening coefficient
$A_{K}/E(B-V)=0.38$ (Savage \& Mathis 1979). 
Fig. \ref{beta} shows the bestfit of eq. (\ref{eq1}). The resulting metallicity
slope turns out to be $\beta=0.08\pm0.11$ in full agreement with the values
estimated in the past empirical studies listed in Table 2. 
The large uncertainty on the coefficient $\beta$ reflects the sparse number of calibrator GCs. However, although such an uncertainty does not allow a conclusive sentence, our analysis confirms 
the discrepancy between the 
observed dependence of the PL$_K$ relation on metallicity and that predicted by
theoretical analyses.
     
\begin{table}
\label{carr}
 \centering
  \caption{Distances and Reddening of the four distance calibrator GCs (from
  Carretta et al. 2000).}
  \begin{tabular}{@{}lccr@{}}
  \hline
   Name     & [Fe/H] & $(m-M)_{0}$  & E(B-V) \\
 \hline
 M5           & -1.10 & 14.46 & 0.035\\
 M15          & -2.15 & 15.38 & 0.090\\
 M68          & -1.95 & 15.13 & 0.040\\
 M92          & -2.15 & 14.64 & 0.025\\
 \hline
\end{tabular}
\end{table}

\subsection{Zero Point of the PL$_K$ relation ($\gamma$)}
\label{zeropoint}

The approach adopted in Sect. \ref{met} allows to derive also the zero point of the
PL$_K$ relation ($\gamma$) directly by the bestfit of eq. (\ref{eq1}) for the 4
GCs with known distances. However, although the relative distances of the
GCs in the Carretta et al. (2000) sample are homogeneous between them, the zero
point of that distance scale could be affected by
systematic errors. As discussed by Gratton et al. (2003), the distance scale
based on the MS fitting method could actually be $\sim 0.1$ mag fainter 
(i.e. longer) than that estimated by Carretta et al. (2000). 
For this reason we decided to calibrate our PL$_K$ relation using the K
magnitude of the variable RR Lyr, whose trigonometric parallax has been
recently determined with HST (Benedict et al. 2002). 
We adopt for RR Lyr the period P=0.5668054 (Kazarovets, Samus \& Durlevich 
2001), the reddening coefficient $E(B-V)=0.02$ (Benedict et al. 2002), 
the metallicity $[Fe/H]=-1.39$ (Clementini et al. 1995) and the distance 
modulus $(m-M)_{0}=7.08\pm0.11$ (Benedict et al. 2002). 
We use the value $K_{2MASS}=6.52$ 
as the average of two measures taken half a cycle apart (Fernley, Skillen 
\& Burki 1993). Given the relatively small amplitude and with 
the additional help of the templates by Jones, Carney \& Fulbright (1996), 
we assumed this average to be  
reasonably close to the mean magnitude. Some additional uncertainty is given 
by the Blazhko effect that modulates the shape of the light curve with a 
double periodicity of 41 days and 4 years (Detre \& Szeidl, 1973). 
Although the Blazhko modulation does not affect the intrinsic value of the 
mean magnitude, its estimate from two data points may be affected. 
So, in absence of a detailed and accurate K light curve for RR Lyr, we adopt 
the mean value K=6.52 (and hence M$_K$=--0.57), and we associate an error 
of $\pm$0.10 mag rather than the 1$\sigma$ error of $\sim$0.07 mag proposed 
by Fernley et al. (1993). 
  
The location of RR Lyr in the $(M_{K}-\alpha~logP_{F})-[Fe/H]$ plane is 
plotted in Fig. \ref{beta} . As can be noted the absolute (period scaled) 
$M_{K}$ magnitude of RR Lyr is 0.1 mag fainter than the value 
predicted by the bestfit of eq (\ref{eq1}) for the 4 GCs listed in Table
3, in agreement with Gratton et al.'s (2003) considerations. 
Then, by constraining our PL$_K$ relation to fit the absolute K magnitude of 
RR Lyr, we derive the zero point of the calibration, $\gamma=-1.05\pm 0.13$. 
The uncertainty on the zero point is calculated by propagating the error on the parallax of RR Lyr with the resulting rms of the best-fit shown in Fig. \ref{alpha} .

As a result of this procedure, we obtain the 
following PL$_K$ relations, in the Carretta \& Gratton (1997) and 
Zinn \& West (1984) metallicity scales, respectively: 
\begin{equation}
\label{eq4}
 M_{K} =  - 2.38(\pm 0.04)~log~P_{F} + 0.08(\pm 0.11)~[Fe/H]_{CG} \\
\end{equation}
~~~~~~~~ $- 1.05(\pm 0.13)$

\begin{equation}
\label{eq5}
 M_{K} =  - 2.38(\pm 0.04)~log~P_{F} + 0.09(\pm 0.14)~[Fe/H]_{ZW}  \\ 
\end{equation}
~~~~~~~~ $- 1.04(\pm 0.13)$

\section{The Distance to the calibrator GCs and to the LMC}

We used such a relation to derive the distances to the
calibrator GCs listed in Table 1. Distances were derived by
bestfitting the eq. (4) to the RR Lyrae K magnitudes for each GC applying the outliers
rejection procedure described in Sect. \ref{period}.
The derived distance moduli are listed in Table 4. 

As an independent check, we apply this procedure to the sample of RR Lyrae stars 
in the inner regions of the LMC whose metallicities (in the Zinn \& West 1984 scale) 
and K magnitudes were derived by Borissova et al. (2004). The observed K 
magnitudes are corrected by --0.044 mag to convert them into the 2MASS 
photometric system according to Carpenter (2001). 
We adopt the reddening coefficient $E(B-V)=0.11$
(Clementini et al. 2003). The resulting distance modulus to the LMC turns out to
be $(m-M)_{0}=18.54\pm0.15$, in good agreement with the most recent
infrared studies of Cepheid variables in the LMC 
(Persson et al., 2004; Gieren et al., 2005). 

\begin{table}
\label{dist}
 \centering
  \caption{Derived distances to the calibrator GCs.}
  \begin{tabular}{@{}lcr@{}}
  \hline
   Name     & $(m-M)_{0}$ & $\alpha_{i}$\\
 \hline
 M3            & 15.07 & -2.34\\
 M4            & 11.39 & -2.39\\
 M5            & 14.35 & -2.27\\
 M15           & 15.13 & -1.99\\
 M22           & 12.65 & -2.50\\
 M55           & 13.62 & -2.94\\
 M68           & 15.01 & -2.28\\
 M92           & 14.65 & -2.31\\
 M107          & 13.76 & -2.39\\
 IC4499        & 16.35 & -1.70\\
 NGC 3201      & 13.40 & -2.26\\
 NGC 5897      & 15.46 & -1.77\\
 NGC 6362      & 14.44 & -2.61\\
 NGC 6584      & 15.67 & -2.05\\
 $\omega$ Cen  & 13.72 & -2.23\\
 Reticulum     & 18.44 & -2.17\\
 \hline
\end{tabular}
\end{table}

\section{Conclusions}

From an accurate analysis of 538 RR Lyrae variables in 16 GCs using infrared 
(K-band) photometry we derive a PL$_K$ relation based on purely observational 
constraints. The derived
dependences of the K magnitude on period and metallicity are in good agreement
with those estimated by previous empirical studies. We confirm that the
metallicity coefficient is about 2-3 times smaller than that predicted by 
theoretical models, as it was found in all previous empirical analyses. 
The zero point of the calibration has been tied to the trigonometric parallax of
RR Lyr measured with HST by Benedict et al. (2002). 
This calibration has been used to derive the distances to the 16 calibrator GCs 
considered in this analysis. As a further check, this relation 
has been applied to RR Lyrae stars in a few central fields in the LMC yielding a
distance modulus $(m-M)_{0}=18.54\pm0.15$,  
in good agreement with the most recent determinations based on Cepheid 
variables (Persson et al., 2004; Gieren et al., 2005).

\section*{acknowledgements}
This research was supported by the Ministero dell'Istruzione, Universit\`a e Ricerca. We warmly thank Paolo Montegriffo for assistance during catalogs cross-correlation. We also thank Santino Cassisi and the anonymous referee for their helpful comments and suggestions.

\label{lastpage}


\begin{thebibliography}{99}

\bibitem[Benedict et al.(2002)]{B02} Benedict G. F. et
al., 2002, ApJ, 581, 115
\bibitem[Bono et al.(2001)]{B01}  Bono G., Caputo F., Castellani V., Marconi M., 
2001, MNRAS, 326, 1183
\bibitem[Bono et al.(2003)]{B03}  Bono G., Caputo F., Castellani V., Marconi M., Storm
J., Degl'Innocenti S., 2003, MNRAS, 344, 1097
\bibitem[Borissova et al.(2004)]{B04} Borissova J., Minniti D., Rejkuba M., Alves D.,
Cook K. H., Freeman K. C., 2004, A\&A, 423, 97
\bibitem[Butler (2003)]{Bu03} Butler D. J., 2003, A\&A, 405, 981
\bibitem[Caputo et al. (1998)]{cap98} Caputo F., Marconi M., Santolamazza P., 
   1998, MNRAS, 293, 364 
\bibitem[Carney et al.(1992)]{C92} Carney B. W., Storm J., Jones R. V, 1992,
ApJ, 386, 663 
\bibitem[Carpenter (2001)]{Ca01} Carpenter J. M., 2001, AJ, 121, 2851
\bibitem[Carretta \& Gratton (1997)]{CG97} Carretta E., Gratton R. G., 1997, A\&AS, 121, 95
\bibitem[Carretta et al. (2000)]{C00} Carretta E., Gratton R. G., Clementini G., Fusi
Pecci F., 2000, ApJ, 533, 215
\bibitem[Catelan et al. (2004)]{C04}  Catelan M., Pritzl B. J., Smith H., A., 2004,
ApJ Suppl. Ser., 154, 633
\bibitem[Chaboyer (1999)]{C99}  Chaboyer B., 1999 in "Post-Hipparcos cosmic
candles", Kluwer Ac. Pub., A. Heck, F. Caputo eds., 237, 111
\bibitem[Clement et al. (2001)]{C01}  Clement C. et al., 2001, AJ, 122, 2587
\bibitem[Clementini et al. (1995)]{C95} Clementini G., Carretta E., Gratton
R., Merighi R., Mould J. R., McCarthy J. K., 1995, AJ, 110, 2319
\bibitem[Clementini et al. (2003)]{C03} Clementini G., Gratton R., Bragaglia A.,
Carretta E., Di Fabrizio L., Maio M., 2003, AJ, 125, 1309
\bibitem[Dall'Ora et al. (2004)]{D04} Dall'Ora M. et al., 2004, ApJ, 610, 269
\bibitem[Del Principe et al. (2005)]{D05} Del Principe M., Piersimoni A. M., Bono G.,
Di Paola A., Dolci M., Marconi M., 2005, AJ, 109, 2714
\bibitem[Detre \& Szeidl (1973)]{ds73} Detre L., Szeidl B., 1973, IBVS, 764
\bibitem[Fernley et al. (1993)]{F93} Fernley J., Skillen I., Burki G., 1993, A\&AS, 97,
815
\bibitem[Fernley et al. (1998)]{F98} Fernley J., Skillen I., Carney B. W.,
Cacciari C., Janes K., 1998, MNRAS, 293, 61
\bibitem[Ferraro et al. (1999)]{F99} Ferraro F. R., Messineo M., Fusi
 Pecci F., De Palo M. A., Straniero O., Chieffi A., Limongi M., 1999
 AJ, 118, 1738
\bibitem[Ferraro et al. (2000)]{F00} Ferraro F. R., Montegriffo P.,
 Origlia L., Fusi Pecci F., 2000, AJ, 119, 1282
\bibitem[Frolov \& Samus (1998)]{FS98} Frolov M. S., Samus N. N., 1998, AstL, 24, 171
\bibitem[Gieren et al. (2005)]{G05} Gieren W., Storm J. B., Thomas G. III, Fouqu\'e P.,
Pietrzy\'nski G., Kienzle F., 2005, ApJ, 627, 224 
\bibitem[Gratton et al. (2004)]{G04} Gratton R. G., Bragaglia A., Clementini
G., Carretta E., Di Fabrizio L., Maio M., Taribello E., 2004, A\&A, 421, 937
\bibitem[Gratton et al. (2003)]{G03} Gratton R. G., Bragaglia A., Carretta E.,
Clementini G., Desidera S., Grundahl F., Lucatello S., 2003, A\&A, 408, 529
\bibitem[Jones et al. (1988)]{J88} Jones R. V., Carney B. W., Latham D. W.,
1988, ApJ, 332, 206
\bibitem[Jones et al. (1992)]{J92} Jones R. V., Carney B. W., Storm J., Latham D. W.,
1992, ApJ, 386, 646
\bibitem[Jones et al. (1996)]{J96} Jones R. V., Carney B. W., Fulbright J. P.,
1996, PASP, 108, 877
\bibitem[Kazarovets et al. (2001)]{K01} Kazarovets E. V., Samus N. N., Durlevich O. V.,
2001, Inf. Bull. Variable Stars, 5135, 1
\bibitem[Liu \& Janes (1990)]{LJ90} Liu T., Janes K. A., 1990, ApJ, 354, 273   
\bibitem[Longmore et al.(1990)]{L90} Longmore A. J., Dixon R., Skillen I., Jameson R.
F., Fernley J. A., 1990, MNRAS, 247, 684
\bibitem[Nemec, Linnell Nemec \& Lutz (1994)]{N94} Nemec J. M., Linnell Nemec A. F., Lutz T. E., 1994,
AJ, 108, 222
\bibitem[Persson et al.(2004)]{P04} Persson S. E., Madore B. F., Krzemi\'nski W., Freedman W. L., 
Roth M., Murphy D. C., 2004, AJ, 128, 2239
\bibitem[Rey et al.(2000)]{R00} Rey S. C., Lee Y. W., Joo J. M., 
 Walker A., Baird S., 2000, AJ, 119, 1824
\bibitem[Rich et al. (2005)]{R05}  Rich R. M., Corsi C. E., Cacciari C.,
Federici L., Fusi Pecci F., Djorgovski S. G., Freedman W. L., 2005, AJ, 129, 2670
\bibitem[Savage \& Mathis (1979)]{sm79} Savage B. D., Mathis J. S., 1979, ARA\&A 17, 73 
\bibitem[Skillen et al. (1993)]{S93} Skillen I., Fernley J. A., Stobie R. S., 
Jameson R. F., 1993, MNRAS, 265, 301
\bibitem[Skrutskie et al. (2006)]{S06} Skrutskie M. F. et al., AJ, 131, 1163 
\bibitem[Smith (1995)]{S95} Smith H. A., 1990 in "RR Lyrae stars", Cambridge University 
Press
\bibitem[Sollima et al.(2004)]{sol04} Sollima A., Ferraro F. R., Origlia L., Pancino 
E., Bellazzini M., 2004, A\&A, 420, 173 
\bibitem[Storm (2004)]{Sto04} Storm J., 2004, A\&A, 415, 987
\bibitem[Valenti et al.(2004)]{valenti} Valenti E., Ferraro F. R., Perina S., 
Origlia L., 2004, A\&A, 419, 139
\bibitem[Valenti et al.(2004)]{VFO04} Valenti E., Ferraro F. R., 
Origlia L., 2004, MNRAS, 351, 1204
\bibitem[van Albada \& Baker (1971)]{vab71} van Albada T. S., Baker N., 
   1971, ApJ, 169, 311
\bibitem[Zinn \& West (1984)]{ZW84} Zinn R., West M. J., 1984, ApJS, 55, 45

\end{thebibliography}
\end{document}